\pdfoutput=1
\documentclass{JINST}
\usepackage{subfigure}

\title{Position sensitive SiPM detector for Cherenkov applications}

\author{L. Gruber $^{a,b,}$\thanks{Corresponding author.}~, G.S.M. Ahmed $^a$, S.E. Brunner $^{a,b}$, 
  P. B\"uhler $^a$, J. Marton $^a$, K. Suzuki $^a$\\
\llap{$^a$}Stefan Meyer Institute for Subatomic Physics, Austrian Academy of Sciences,\\
  Boltzmanngasse 3, 1090 Vienna, Austria\\
\llap{$^b$}Vienna University of Technology, Faculty of Physics,\\
  Karlsplatz 13, 1040 Vienna, Austria\\
\\
  E-mail: \email{lukas.gruber@oeaw.ac.at}}

\abstract{A prototype of a position sensitive photo-detector with 5.6 $\times$ 5.6 cm$^2$ detection area readout with 64 Hamamatsu MPPCs
  (S10931-100P) with 3 $\times$ 3 mm$^2$ active area each has been built and tested. The photo-sensors are arranged in a 8 $\times$ 8 array
  with a quadratic mirror light guide on top. The module is currently readout by in-house developed preamplifier boards but employing
  existing ASIC chips optimized for SiPM readout is also planned. Such a device is one of the candidates to be used for photon
  detection in the PANDA DIRC detectors.}


\begin{document}

\section{Introduction}

Silicon Photomultipliers (SiPMs) are extremely versatile photo-sensors and can be used in many fields ranging from astrophysics, particle and
nuclear physics to medical imaging.

Recently, we have built a position sensitive photo-detector based on an array of SiPMs which could be used for photon detection in the PANDA
experiment at the FAIR facility in Darmstadt, Germany. For charged particle identification in the momentum range of 0.5 GeV/c to 4.5 GeV/c, two
DIRC detectors are foreseen. The barrel DIRC detector \cite{barrelDIRC} has a total detection area of a few m$^2$ which will be covered by about
10 000 photo-sensors. The DIRC detector must be able to detect Cherenkov light at very low intensities with an angular resolution of 2 - 2.5 mrad.
Microchannel plate PMTs (MCP-PMTs) are considered as an option for photon detection in the PANDA DIRCs. However, the lifetime is still not
sufficient for the expected photon rates \cite{MCPstudy} and therefore we are studying SiPMs as an alternative to MCP-PMTs.

\section{Position sensitive Cherenkov detector}

\subsection{Light concentrator}

\begin{figure}[t] 
  \centering
  \includegraphics[width=0.9\textwidth]{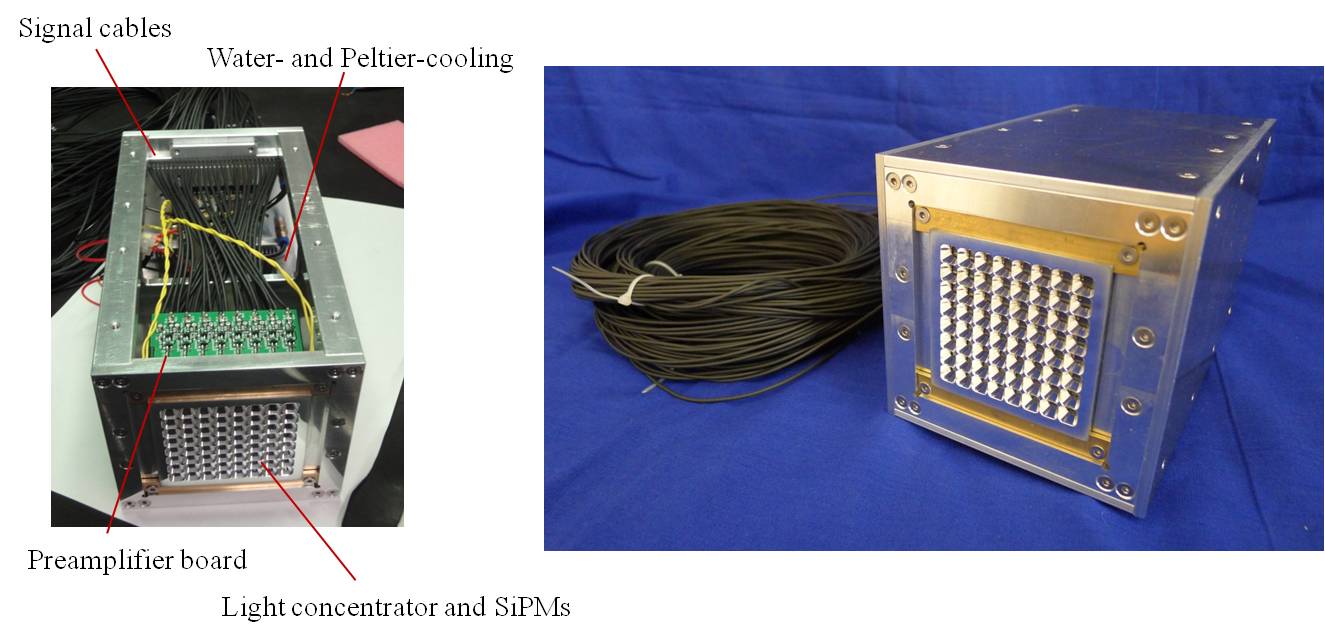}
  \caption{Pictures of the position sensitive SiPM array. The figure on the left shows a view inside the detector. The SiPMs are read out from
  the back by four preamplifier boards. The temperature is stabilized by water- and Peltier-cooling. The light guide matrix on top of the
  SiPMs enlarges the detection area of the module. On the right hand side the detector is closed.}
  \label{fig:prototype}
\end{figure}

In order to increase the number of incident photons on the active area of the sensor, the idea of using an array of suitable light guides on
top of the photo-sensor has been studied. Such a light concentrator leads to increased geometric acceptance and increased signal to noise ratio,
since the dark count rate is not affected by the light guides. The light concentrator consists of 64 regularly arranged pyramid-shaped funnels
with quadratic (round edges) entrance windows of 7 $\times$ 7 mm$^2$ and exit apertures of 3 $\times$ 3 mm$^2$, respectively, and thus increases
the geometric detection area by a factor of (7/3)$^2 \times \epsilon_{geo} \approx$ 5.1, where $\epsilon_{geo} = 0.93$ is the geometric fill
factor of the light concentrator. The funnel height is 4.5 mm. The light guide array is made out of brass and the funnels were produced by
electro-erosion. Two modules of the light concentrator with different coatings (Aluminum and Chromium) were produced and tested.
Figure \ref{fig:prototype} shows a picture of the light concentrator mounted on top of the SiPM matrix.

\subsection{Prototype detector}

The prototype of a position sensitive photo-detector consists of 64 SiPMs (Hamamatsu MPPC S10931-100P) with 3 $\times$ 3 mm$^2$ active area
and 100 $\times$ 100 $\mu$m$^2$ pixel size, arranged in a 8 $\times$ 8 array. The SiPM array is combined with a light concentrator on top.
Each photo-sensor is read out separately. An individual bias supplier for each SiPM ensures that the gains of the sensors can be adjusted to be the
same. The four preamplifier boards, developed at SMI, consist of 16 preamplifiers each and provide sufficient signal amplification with a gain
of around 5 and reasonably fast shaping time ($\sim$ 2 ns). A picture of the prototype detector is shown in Figure \ref{fig:prototype}.

\section{Efficiency measurements and simulations}

\subsection{Measurements of the light concentrator efficiency}

In order to estimate the collection efficiency of the light concentrator, the SiPM array, with the light concentrator on top, is
scanned in two dimensions with a blue laser (407 nm). The beam spot of about 1 mm diameter is moved in steps of 500 $\mu$m and the
average output pulse height is recorded with an oscilloscope. The expected incident angle is $\theta = 0 \pm 4^\circ$. The measurements
are done inside a dark box. Since it is known that the key parameters of SiPMs show a strong temperature dependence \cite{SiPMstudy, timingstudy},
the whole setup is kept stable at 15 $^\circ$C, using water- and Peltier-cooling.

The detection efficiency of a single funnel can be written as $\epsilon_{detect} = \epsilon_{col} \times \epsilon_{PDE}$, where
$\epsilon_{col} = n_{d}/N_{phot}$ is the collection efficiency of the light concentrator, with $n_{d}$ being the number of photons reaching
the exit aperture and $N_{phot}$ being the total number of photons hitting the entrance aperture, and $\epsilon_{PDE}$ is the photon detection
efficiency of the photo-sensor.

\begin{figure}[t] 
  \centering
  \subfigure{
  \includegraphics[width=0.48\textwidth]{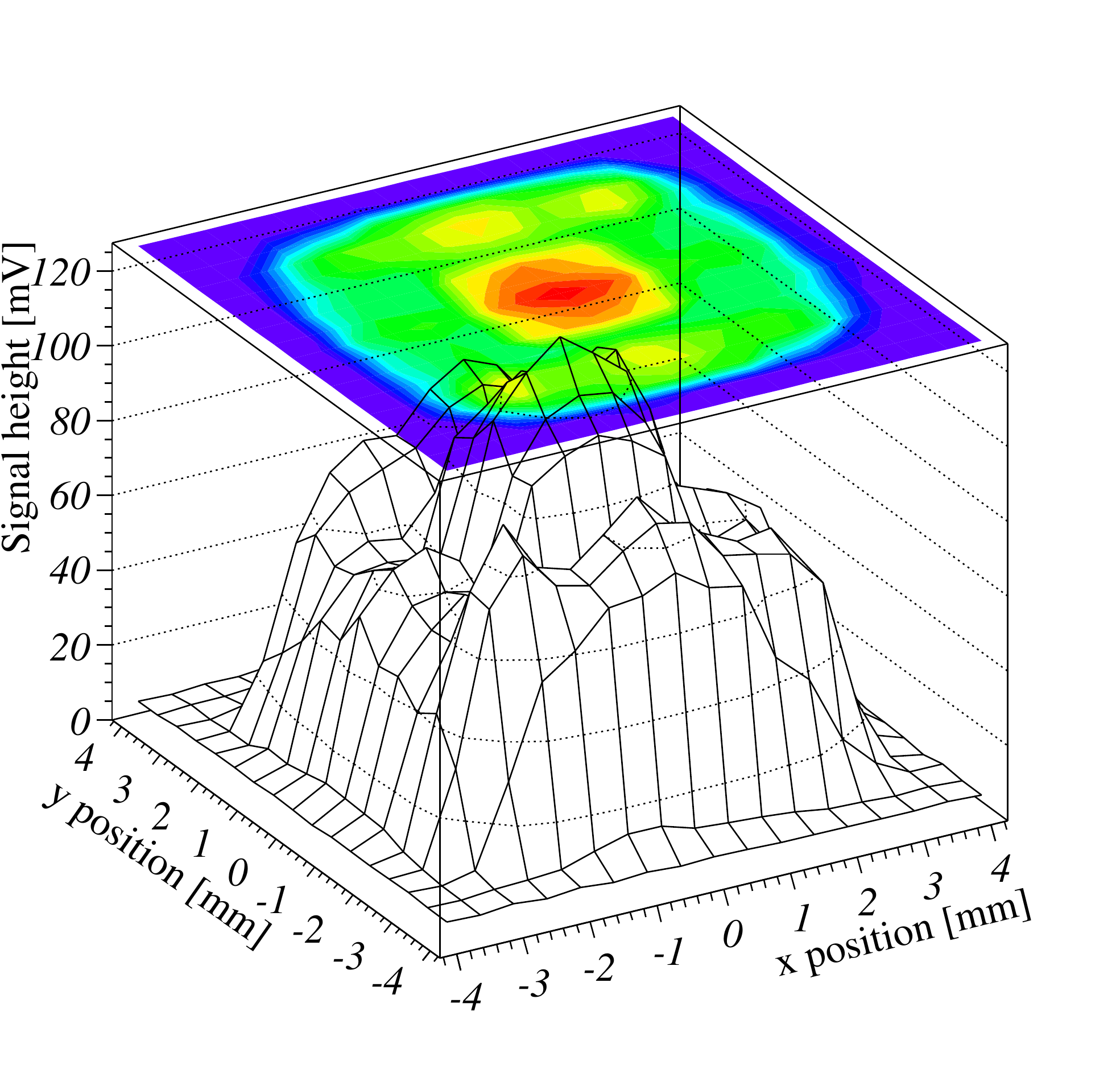}
  }   
  \subfigure{
  \includegraphics[width=0.48\textwidth]{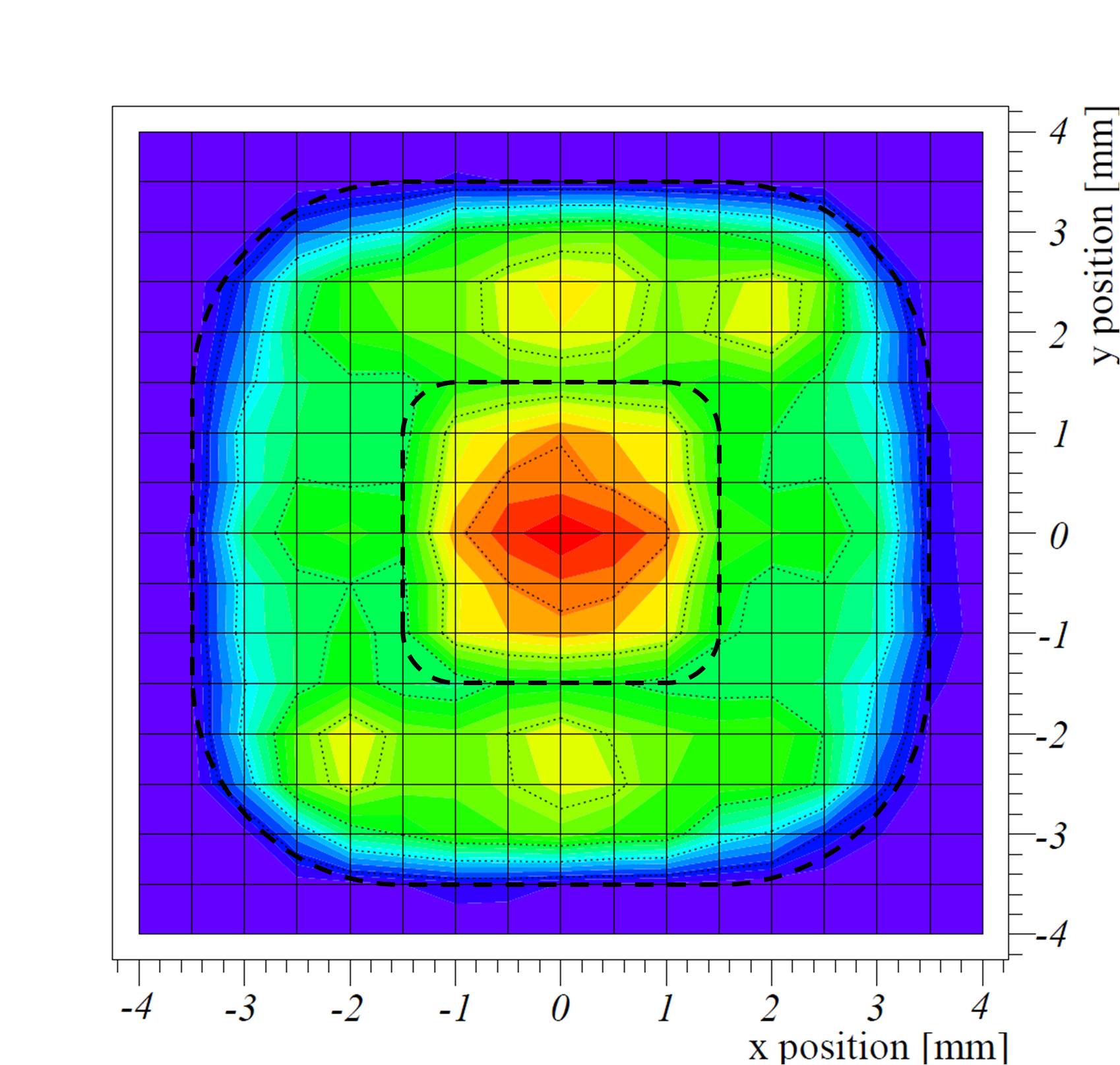}
  }
  \caption{The plots show the results from a scan of one cell of the SiPM array with a laser spot of about 1 mm diameter in steps of 500 $\mu$m
  in 3D- (left) and top-view (right). The chromium-plated light concentrator is used. The dimensions of the entrance- and exit aperture are
  indicated in the right plot.}
  \label{fig:ScanXY}
\end{figure}

\begin{figure}[t]
  \centering
  \subfigure{
  \includegraphics[width=0.43\textwidth]{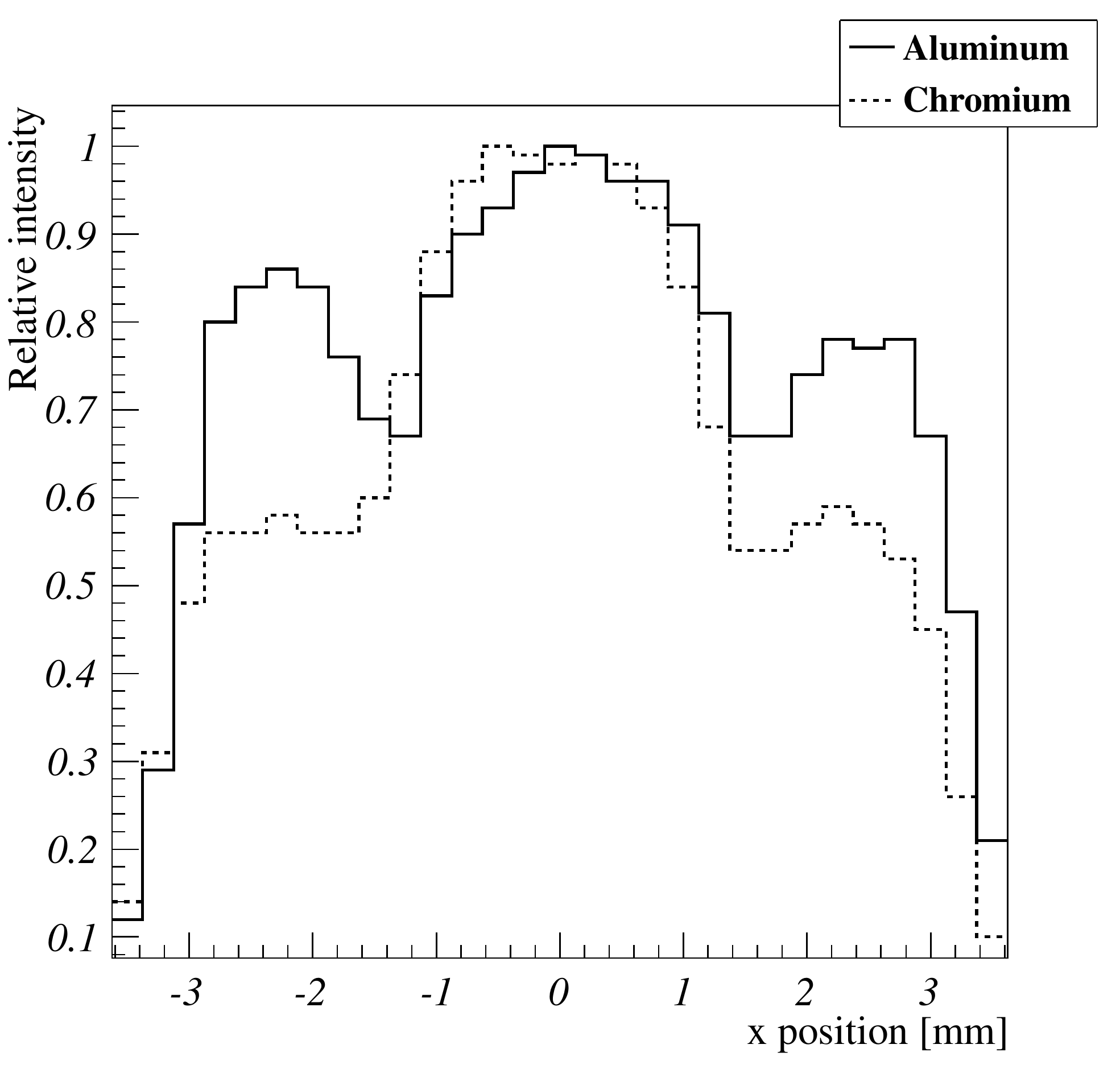}
  }   
  \subfigure{
  \includegraphics[width=0.43\textwidth]{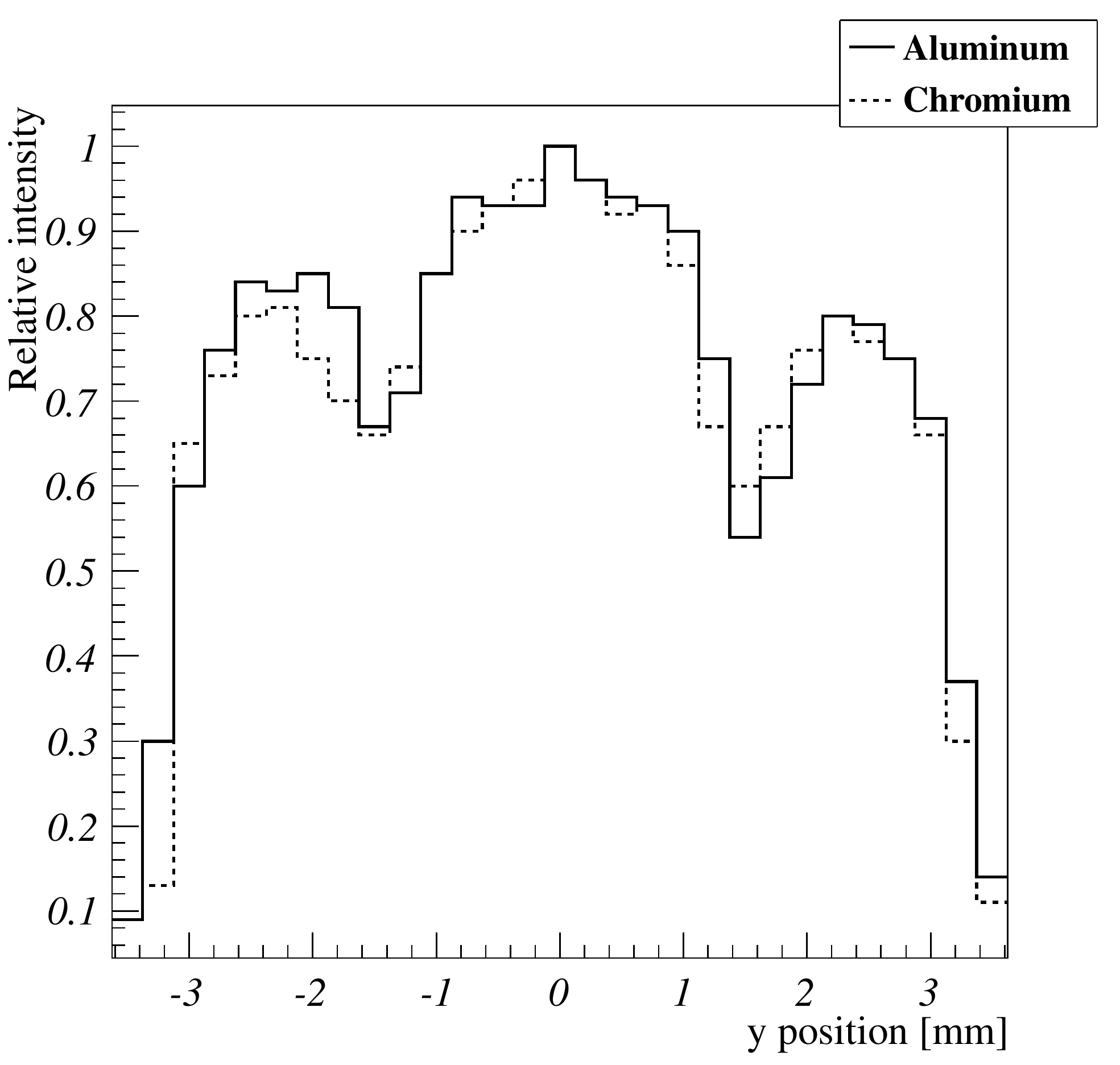}
  }
  \caption{The plots show a comparison between the Aluminum- and Chromium-plated light guide. One cell is scanned in x-direction (left) at
  fixed y-position ($y = 0$) and in y-direction (right) at fixed x-position ($x = 0$). On average, the Aluminum-plated light concentrator
  performs better than the Chromium-plated one.}
  \label{fig:ScanE5comp}
\end{figure}

Figure \ref{fig:ScanXY} shows a measurement done for one cell (one funnel plus SiPM) of the array, using the Chromium-plated light
concentrator. One can identify the edges of the entrance and exit aperture at $x = y = \pm$ 3.5 mm and $x = y = \pm$ 1.5 mm, respectively,
and the SiPM centered at $x = y = 0$. The use of the light concentrator clearly helps to increase the detection area of the module. The partly
inhomogeneous surface and xy-asymmetry of the light concentrator is supposed to originate from a non-uniform coating quality. The fact that
the profiles of other cells reveal very similar features strengthens this suspicion. Besides, there are less efficient areas at $\pm$ 1.5 mm.
This becomes more obvious when looking at Figure \ref{fig:ScanE5comp}. The plots show the profiles of a scan for the same cell in
x- and y-direction in steps of 250 $\mu$m, comparing the Chromium- and Aluminum-plated light concentrator. The less efficient bands at a
position of $\pm$ 1.5 mm are likely due to defects, oxidation or bad coating at the edges of the funnel or due to none perfect matching
between light guide and SiPM, but could also be a feature of the photo-sensor. This needs further investigation. Nevertheless, the fact that
the Aluminum coated light concentrator shows a symmetric behavior in x- and y-direction confirms the previous assumption of a non-uniform
coating quality in case of the Chromium-plated light guides. Other cells have been tested and provided comparable results.

From the measurements we find an average light collection efficiency of $\epsilon_{col} = 57\%$ for the Chromium-plated light concentrator,
which shows that it is working quite efficiently. In order to compare with simulation, the data from the above measurments are normalized to
the maximum signal height, so that the photon detection efficieny $\epsilon_{PDE}$, which is not considered in the simulation, is canceled out.

\subsection{Simulations of the light concentrator efficiency}

A series of Monte Carlo simulations using a self-developed code was carried out in order to evaluate the collection efficiency $\epsilon_{col}$
of the light concentrator \cite{simulations}. Assuming a reflection coefficient of 0.55 (Chromium at 400 nm \cite{refrindex}) and a reasonable
surface smoothness, the simulations show an average light collection efficiency of around $52\%$ at $\theta = 0^\circ$,
where $\theta$ is the angle relative to the aperture normal, which is in very good agreement with the measured value of $57\%$.

\section{Conclusions and outlook}

A prototype of a position sensitive photo-detector with SiPM readout has been built and tested. Measurements were done in our laser laboratory
and recently also at the T9 test beam at CERN in Geneva, Switzerland. The data recorded in ten days beam time are currently under investigation.
It was shown that the light concentrator clearly increases the acceptance of the module and that the efficiency measurements are in good
agreement with the simulations. The Aluminum-plated light guide is the preferred solution for future measurements. Further laser tests with
different incident angles and smaller laser spots in the range of the MPPC pixel size and below are planned. In a next step it's also foreseen
to use existing ASIC chips for the SiPM readout.

\acknowledgments

This work is partly supported by the EU Project HadronPhysics 2 (project 227431).

\end{document}